# "Tom" pet robot applied to urban autism

Xingqian Li[1], Chenwei Lou[1], Jian Zhao[1], HuaPeng Wei[1], Hongwei Zhao[1]

*Abstract*—With the fast development of network information technology, more and more people are immersed in the virtual community environment brought by the network, ignoring the social interaction in real life. The consequent urban autism problem has become more and more serious. This article focuses on "promoting offline communication between people " and "eliminating loneliness through emotional communication between pet robots and breeders" to solve this problem, and has developed a design called "Tom". "Tom" is a smart pet robot with a pet robot-based social mechanism called "Tom-Talker". The main contribution of this paper is to propose a social mechanism called "Tom-Talker" that encourages users to socialize offline .And "Tom-Talker" also has a corresponding reward mechanism and a friend recommendation algorithm. It also proposes a pet robot named "Tom" with an emotional interaction algorithm to recognize users' emotions, simulate animal emotions and communicate emotionally with users. This paper designs experiments and analyzes the results. The results show that our pet robots have a good effect on solving urban autism problems.

## I. Introduction

In recent years, due to the increase of online virtual communities, people have become accustomed to meeting new friends through the Internet. The intimate contact between people has been replaced by network links. Urban autism has become increasingly popular and aggravated [1]. People are apt to have a lot of online friends, but in real life, probably does not have any good friends[2]. The interpersonal relationship brought by social networking sites is becoming superficial and low quality[3].

In the past, people can try to keep pets to eliminate their loneliness. Pets have always been an important partner of people and played an important role in helping people keep a good mood [4]. And the accompanying "pets social" as a common vertical social way [5] also played an important role in eliminating people's loneliness and expanding people's social circle [6]. But because of the cleaning and disease problems caused by pets, fewer people are keeping pets. This important way to eliminate people's loneliness is gradually losing its role.

At the same time, with the development of computer technology, robots have gradually become close partners in human social life and become an indispensable part of human life. With the development of service robot technology and the continuous decline in the cost of robot production, intelligent robots will enter thousands of households and become good helpers for people's life [7], work and study. Most pet robot designers today tend to pay more attention to the entertainment functions of robots. The animal emotion simulation model is often lacking inside their products, and the emotion changes of the breeders cannot be felt. There is a serious emotional barrier between these pets and the breeders [8]. So they can't play a companion role. In addition, designers tend to focus only on pet companionship, ignoring the added value of "pets social" [9]. We believe that designing a pet robot that can communicate emotionally with users without cleaning and disease problems, and is attached with a suitable pet social mechanism will have practical significance and value for solving urban autism problems.

In view of the above background, this paper designs a pet robot-based social mechanism called "Tom-Talker" and a pet robot named "Tom". The "Tom-Talker" social mechanism is a new online social mechanism consisting of users, stores and a social platform, which can match users to friends and issue offline tasks to motivate users to socialize offline. "Tom" pet robot can recognize the user's emotions, simulate the emotions of real animals, and communicate emotionally with the users to ease their loneliness. This article adopts the design ideas of the two paths of "breeders and pets" and "breeders and breeders" to solve the increasingly serious social psychological problems. The architecture of "Tom-Talker" social mechanism is shown in Fig. 1.

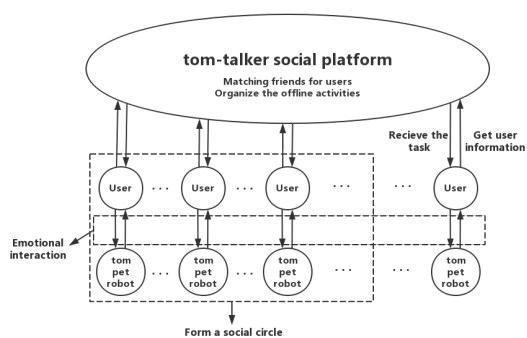

Fig. 1. "Tom-Talker" social mechanism architecture.

## II. Hardware Composition of The "Tom" Pet Robot

The "Tom" pet robot uses a Raspberry Pi as the central processor and a Arduino as the auxiliary control unit. The Raspberry Pi controls interactive modules such as cameras, display screen, and microphones, and communication

Email: Xingqian Li
Lixingqiancs@gmail.com

[1] College of Computer Science and Technology, Jilin University, Changchun 130012, China

modules such as Bluetooth, NFC, and WIFI. Arduino is mainly responsible for data transmission and controls of various environmental information acquisition sensors, such as smoke sensors, light sensors, temperature and humidity sensors, ultrasonic sensors, and so on. Arduino sends the collected information to the Raspberry Pi to operate and control the entire system. The hardware design of "Tom" is shown in Fig. 2.

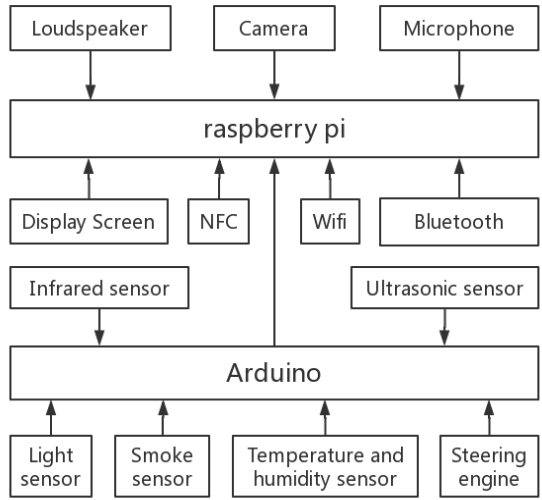

Fig. 2. "Tom-Talker" social mechanism architecture.

Through the Bluetooth module and WIFI module of the Raspberry Pi, users can use the mobile APP to control the behaviors of the robot, and view the emotional changes inside the robot. Users can feed their virtual props to the robot through the mobile phone's NFC module. The camera is used to help the robot recognize the user's emotions. Temperature and humidity sensors, light sensors, etc. enable pets to sense the external environment. The comfort of the external environment and users' emotions can affect the emotion of the robot. The robot can interact with users through the mobile app, display screen, loudspeaker, etc.

III. "TOM-TALKER" SOCIAL MECHANISM

*A. "Tom-Talker" Social Platform Architecture*

Vertical socialization is simply a term for the communication of a group of people with similar interests. Compared with general social networking, vertical social communication has targeted information, which can meet the needs of users in certain aspects, and also has certain commercial value.

"Pets social" is a specific form of vertical social field, which is in line with vertical social characteristics and is a long tail component in the long tail model of social field [10]. For these features, we designed a social platform system with a friend recommendation algorithm and a reward mechanism. It is mainly composed of three parts: social platform, users, and physical stores. The relationship between the three is shown in Fig. 3.

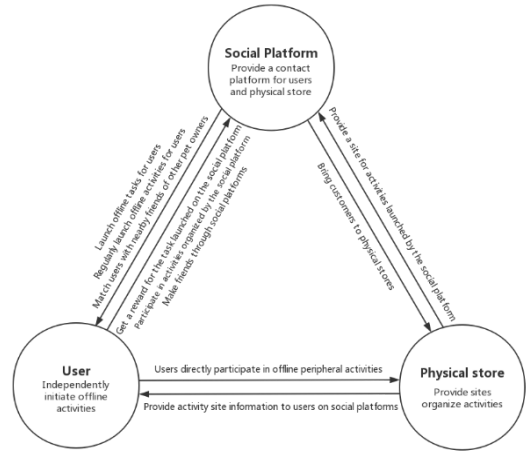

Fig. 3. Social platform, user and store relationship diagram.

The activities of the social platform mainly include:
1. Match the other pet breeders nearby to the user.
2. Regularly organize large offline events.
3. Provide a platform for users and physical stores.
4. Publish offline social tasks.

The user's activities mainly include:
1. Receive the tasks posted on the social platform and get rewards
2. Meet friends on the social platform
3. Independently initiate offline collective activities
4. Participate in large events organized by the platform

The activities of the store mainly include:
1. Provide event information for users on the social platform
2. Provide users with information of rentable venues on the social platform

The social platform recommends friends for users and posts tasks through a friend recommendation algorithm and users must meet their friends offline to complete the task and get rewards. Users can also reflect to the social platform the activities they are more willing to participate in, so that the platform can organize better activities.

*B. "Tom-Talker" Social Incentives*

A reward mechanism [11] usually rewards some of the user's participation behavior, thereby motivating them to participate in more activities.

Those rewards usually include the following types:
1. Complete-mission rewards: rewards that users receive after completing a specific mission.
2. Surprise-collective rewards: rewards that are usually found in some collection or exploration games. Players can earn certain rewards by collecting and exploring.
3. Achievement-based rewards: rewards that are widely found in social platforms where users can show others what they have achieved and gain a sense of spiritual satisfaction.
4. Offline rewards: rewards such as some physical items.

In the long-term research, the reward mechanism has always been regarded as an important external factor to stimulate user participation. Whether a reward mechanism is excellent depends mainly on the value and entertainment of rewards that it can bring to users [12]. Complete-mission rewards is the most basic type. Among the many reward mechanisms that are currently available, it is often the main way for users to get rewards. But a predictable, task completion-related reward is often less exciting than an unpredictable, accidental reward. Therefore, in the reward mechanism designed in this paper, complete-mission rewards and surprise-collective rewards are introduced and integrated. In addition, user viscosity is often an important factor in determining whether a reward mechanism is excellent. An achievement-based reward with a long completion period and a strong sense of accomplishment tends to have a better retention of user viscosity.

The relationship between the various rewards in the system is shown in Fig. 4.

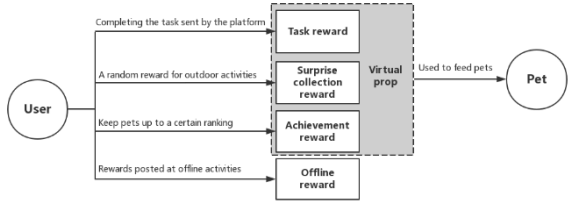

Fig. 4. The relationship between various rewards in the system.

The reward mechanism designed in this paper introduces two kinds of rewards: pet virtual props rewards and physical rewards. The pet virtual props are virtual items such as rations necessary for the breeders to feed their pets, while the physical rewards are some offline items. Rewards available to users include complete-mission rewards, surprise-collective rewards, achievement-based rewards, and offline rewards. The complete-mission rewards are mainly the rewards obtained by the breeder to complete the tasks sent by the platform, mainly composed of virtual props. Surprise-collective rewards are rewards that breeders receive randomly when they are outdoors, including virtual props rewards and physical rewards. Achievement-based rewards are to give the breeder a spiritual satisfaction by showing the breeder's task completion and pet's growth on the social platform. The offline rewards are mainly some physical items given in the large offline events organized by the platform.

Our reward mechanism differs from other reward mechanisms in that it prefers users to go out and participate in social activities. Therefore, the user's participation in social activities will be the main factor determining how much reward they get. In the question of determining how the reward should be distributed, we draw on the idea of graph theory. A weighted undirected graph is constructed with each user in the social system as a node and weighted by the number of finished tasked between the two users. The weight of the connected sides of a user node and the number of offline collective activities that he participates in will be used to calculate how many rewards he receives.

$$R = \alpha * +(\omega_1 + \omega_2 \cdots) + (1-\alpha) * w \quad (1)$$

$R$ represents the total reward obtained by a user, $\omega 1 \sim \omega n$ represents the weight of each edge of the user's node, and $W$ represents how many collective activities the user have participated in. $\alpha$ is a number used to balance the weight of collective activities and user's self-social activities, with a value of $0\sim1$. It is dynamically changing according to the strategy of the social platform.

The value of $\omega$ of each side is calculated according to the numbers of offline tasks users have finished, and the two are positively correlated. We hope that its growth rate will gradually increase in the initial stage, so as to encourage users to participate more in the activity, and after reaching a certain threshold, the growth rate will gradually slow down, so as to prevent users from only communicating with a few people and no longer expanding their social circles. We use the Logical Growth Model to represent this process.

$$\omega_i = q_1/(1 + e^{-p_1(m_1-c_1)}) \quad (2)$$

Where $\omega_i$ is the the weight of the edge with the user $i$, $q_1$ and $p_1$ are the adjustment coefficients, $m_1$ is the number of finished offline tasks with the user $i$, and $c_1$ is the threshold. The variation curve of $\omega_i - m_i$ is shown in Fig. 5.

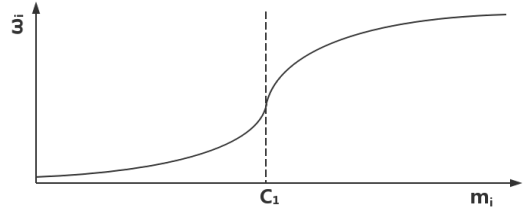

Fig. 5. $\omega_i - m_i$ curve.

### C. "Tom-Talker" Friend Recommendation Algorithm

Friendship is the most basic part of our social system. Whether we can match the right friends to our users determines the quality of our social platform. A friend recommendation refers to recommending a new friend according to factors such as the user's interests and hobbies, and the resulting recommendation algorithm is called link prediction. How to make recommendations more accurately and reasonably will be a problem worth studying.

Current mainstream recommendation algorithms include:
• Content based matching. Recommend friends with similar basic information to users, such as the same cell or the same school, the users often know each other before.
• Hobbies based matching. Recommend friends with similar hobbies to users.
• Social network based matching. Recommend other users in the same social circle through the graph theory algorithm

These three algorithms have their own advantages and disadvantages, but they are all not suitable for our social system. The main purpose of our social system is to recommend users who can conduct offline activities, so the geographical location information will account for a large proportion, and a simple hobby-based matching can not meet our requirements. Matching based on content or social networks tends to recommend users who are already in their social circle, and do not have the ability to expand the social circle. Therefore, in our algorithm, we will comprehensively consider the geographical location, hobbies and social network factors.

The collaborative filtering algorithm [13] is the earliest and more famous recommendation algorithm. Its main function is prediction and recommendation. The algorithm discovers the user's preferences by mining the user's historical behavior data, and classifies the users based on different preferences. The collaborative filtering recommendation algorithm is divided into two categories, namely user-based collaborative filtering algorithm (user-based collaboratIve filtering) and item-based collaborative filtering algorithm (item-based collaborative filtering).

The main steps of the user-based collaborative filtering algorithm include:

1 Find the law from the user's behavior and preferences, and give recommendations based on this.

2 Analyse users' behavior and obtain users' preferences, similar users are calculated according to their preferences, and then recommended based on similar users.

The calculation of similar users mainly includes the following steps:

• Calculate similarity

For the calculation of similarity, several existing basic methods are based on vectors, that is, the distance between two vectors is calculated. The closer the distance is, the more similar the similarity is. After constructing the user-item two-dimensional matrix, a user's preference for all items is used as a vector to calculate the similarity between users.

• Calculation of similar neighbors

After the similarity is calculated, several other users with the highest similarity are selected for recommendation. Fig.6

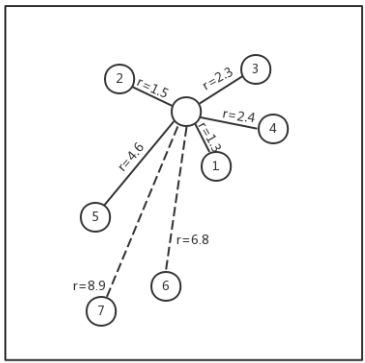

Fig. 6. Recommendation of similar neighbors.

The collaborative filtering algorithm calculates the similarity between users by scoring the products, ignoring the characteristics of the users themselves. The recommendation algorithm designed in this paper improves it. Firstly, we add users' preference for the items together with some basic information of them as parameters to calculate the similarity. Secondly, We introduce distance factors to our model. Among the recommended friends, there are many friends whose similarity is too low, or the geographical distance is too far. This kind of recommendation doesn't have practical significance. We have added two thresholds here. A user is not recommended when his or her similarity is below the threshold, and is not recommended when his or her geographical distance is too far. This will avoid some invalid recommendations. The new recommendation process is shown in Fig. 7.

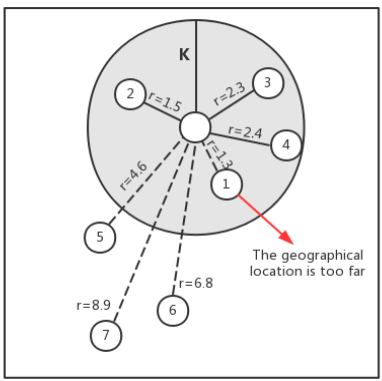

Fig. 7. Improved recommendation model.

According to the recommendation algorithm described above, a user can quickly get some friends with the same interests and a close distance. But this kind of recommendation ignores other friends hidden in the social network. It pays more attention to interests and basic information, and ignores the importance of social network. This will lead to a large number of social circles gathered by interest, and users are still isolated into several parts. Therefore, in our recommendation algorithm, we will pay more attention to the value of social networks when social circles tend to be stable in an area, and recommend friends between social circles with the same intersection.

At present, the predecessors proposed a recommendation algorithm based on the public intersection circle of friends [14]. It is believed that if two users have more overlapping social circles, more overlapping friends, and more closely connected public neighbors, they are more inclined to establish a friend relationship. In this paper, we have modified its recommendation algorithm to take into account the similarity between two non-adjacent users and their public neighbors, as well as the similarity between the two users, to prevent recommendations from public neighbors with low similarity to meet our requirements better.

For two users $u$, $v$ with public neighbors but don't know each other, their network-based recommendation value $S$ can be calculated as in equation (3).

$$S = \alpha * \sum_{i=1}^{n}(n_1 * (m_i + 1)) + (1-\alpha) * \left(\sum_{i=1}^{n_i}\sum_{j=1}^{n_j} e_{ij} + E\right) \quad (3)$$

Where $S$ is the recommendation index and n is the number of independent subgraphs in the common neighbor subgraph of $u$ and $v$. $n_i$ represents the number of points in the independent subgraph $i$ in the common neighbor subgraph of $u$ and $v$, and $m_i$ represents the number of edges in the independent subgraph $i$ in the common neighbor subgraph of $u$ and $v$. $e_{ij}$ represents the average of the similarities between points $j$ and $u$, v in the independent subgraph $i$. $E$ represents the similarity between $u$ and $v$, and $\alpha$ is an adjustment coefficient ranging from $0$ to $1$.

IV. "TOM" PET ROBOT EMOTIONAL INTERACTION ALGORITHM

A. User Facial Emotion Recognition

The Convolutional Neural Network (CNN) [15] is an efficient identification method developed in recent years. In 2012, Krizhevsky et al. used the extended depth CNN to achieve the best classification effect in the ImageNet large-scale visual recognition challenge competition [16], which made CNN more and more valued by researchers. CNN can effectively reduce the complexity of a network, reduce the number of training parameters, and are also easy to train and optimize. Based on these superior features, it outperforms standard fully-connected neural networks in a variety of signal and information processing tasks [17], and performs well for large image processing. General convolutional neural network processing is shown in Fig. 8

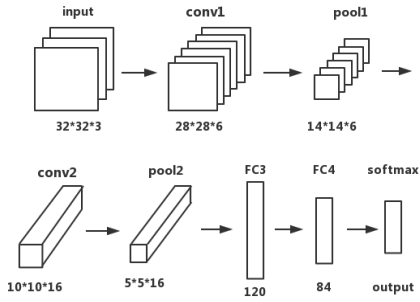

Fig. 8. General convolutional neural network processing.

Residual Network (ResNet) [18] is a variant neural network. In ordinary deep network models, as the depth of the network increases, the accuracy begins to saturate and then declines rapidly. If the layers behind the deep network are all identical mappings, the model degenerates into a shallow network. In order to solve the learning identity mapping function, the network can be designed as:

$$H(x) = F(x) + x \tag{4}$$

At this point, the problem is converted to learning a residual function:

$$F(x) = H(x) - x \tag{5}$$

As long as $F(x) = 0$, an identity map $H(x) = x$ can be constructed, making the fitting easier. The residual network can solve the problem of degraded performance of deep convolutional neural networks without increasing the number of parameters and computational complexity, so that the network can be very deep. The widespread use of residual networks has pushed the performance of computer vision tasks to new heights. A structure of the residual network is shown in Fig.9[19].

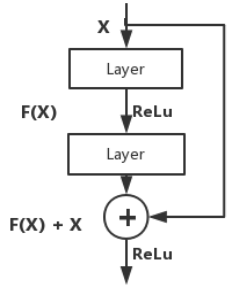

Fig. 9. A structure of the residual network.

Depthwise separable convolution [20] is a separable convolution method commonly used in neural networks. The use of depth separable convolution reduces the required parameters over ordinary convolution. What is important is that the depth separable convolution changes the conventional convolution operation into consideration of both the channel and the area. The convolution first considers only the area, and then considers the channel to achieve separation of the channel and the area. The process of it is shown in Fig. 10 [21].

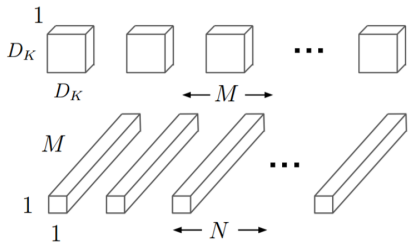

Fig. 10. Depth separable convolution process.

This article uses the dataset provided by the Kaggle Facial Expression Recognition Challenge. There are 7 facial expression categories in this data set: {anger, disgust, fear, happy, sad, surprise, neutral}. There are a total of 35,887 48px×48 px grayscale face maps in this data sets, the samples of which are shown in Fig. 11.

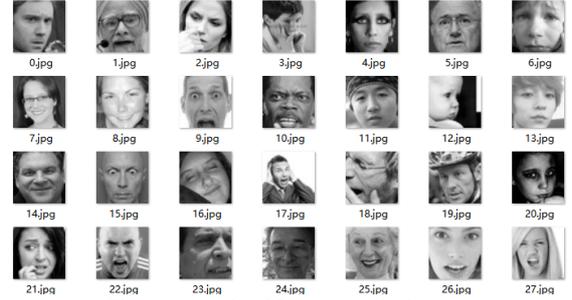

Fig. 11. Sample of face grayscale data.

Before training, in order to solve the problem of too few training sets, we used image augmentation techniques such as random rotation, translation, scaling, and flip to expand our data set. In this way, our model will never see two identical images twice. This helps prevent overfitting and makes the model better generalized.

CNN has developed many mature network models to date. Such as LeNet-5 [22], AlexNet [9], VGG-16 [23] and so on. The Inception v3 [24] model calculates an input by several different feature extraction methods simultaneously, and the network itself learns what parameters it needs; while the Xception model [25], as an improvement of Inception v3, introduces the deep separable convolution, which improves the model's effect without substantially increasing the complexity of the network.

The model we used draws on the idea of the Xception model. Based on the Xception model, we introduce the BatchNormlization[26] layer. The function of this layer is to make the average value of the output data close to 0, and the standard deviation is close to 1, and then correct it. The corrected data is sent to the activation function, namely:

$$\hat{y}^{(k)} = \frac{y^{(k)} - E[y^{(k)}]}{\sqrt{D[y^{(k)}]}} \tag{6}$$

$$y^{(k)} = \gamma^{(k)} \hat{y}^{(k)} + \beta^{(k)} \tag{7}$$

The $\gamma$ and $\beta$ parameters in equation (7) need to be learned. The advantage of using the BatchNormlization operation is that a large learning rate can be used in training, which greatly increases the training speed and greatly speeds up the convergence process.

Considering that the Raspberry Pi has limited computing power, it is necessary to run some other necessary programs while running the emotion recognition program. In order not to affect the recognition speed of the model and the fluency of the Raspberry Pi system, it is necessary to modify the original model.

Fig. 12 shows the specific flow of the model. The model removes the last fully connected layer and can reduce the amount of parameters by using depth separable convolution. The model consists of four residual depth separable convolution blocks, each of which is followed by a BatchNormlization operation and then sent to the ReLu activation function. Finally, the global average pooling layer and the softmax activation function are used to generate the prediction.

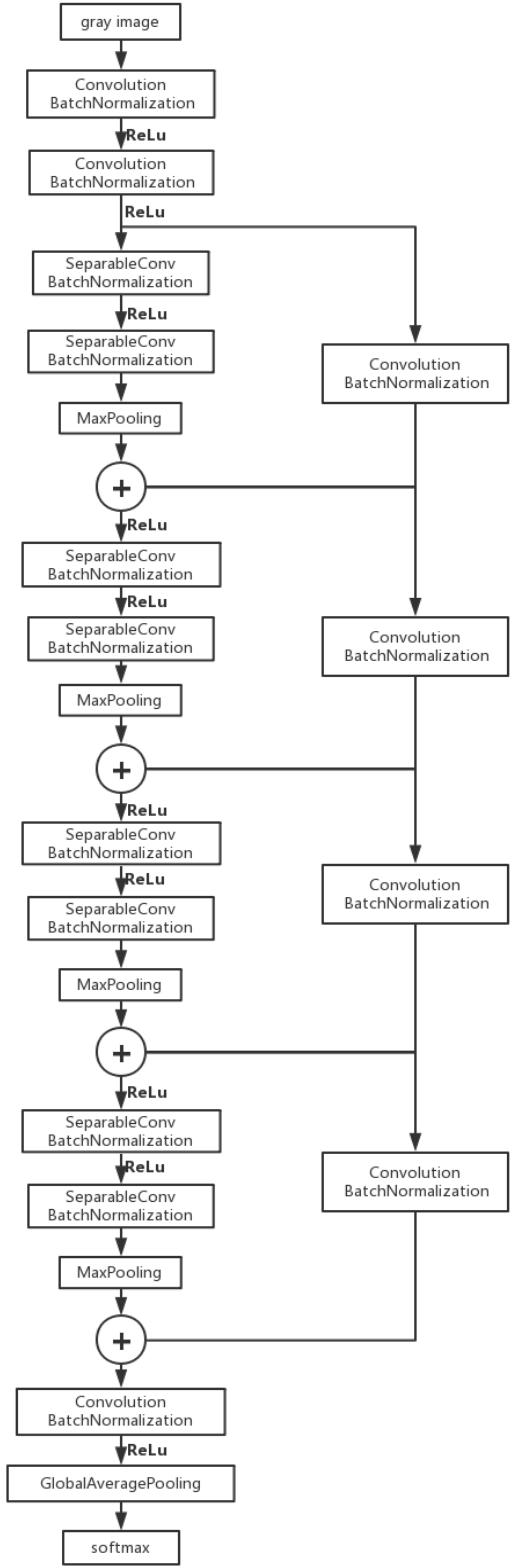

Fig. 12. Our facial emotion recognition model.

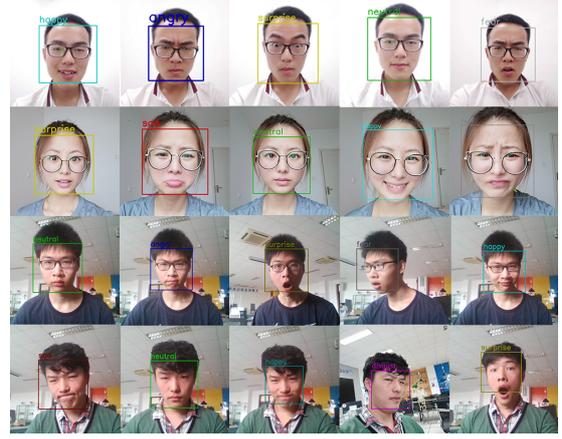

Fig. 13. Facial emotion recognition effect.

The trained model is used to perform real-time facial emotion recognition, convert the picture in the video stream into a grayscale image, detect the location of the face, separate the face and perform emotion detection. The best fit of the emotion is taken as the final fitting result, and the degree of fitness of the emotion is taken as the intensity of the emotion. Fig. 13 shows some test samples, in which the vividness of the color of the box indicates the intensity of the emotion.

### B. Pet robot emotion simulation

There are many algorithms for robotic emotion simulation, but most of them are more like humans, but this obviously does not apply to pet robots. We hope that our pet robot has a more animal-like emotional simulation algorithm, and hopes that it can respond according to the emotion of the breeder to achieve the purpose of eliminating loneliness. Here we define the seven basic emotions of the robot {anger, disgust, fear, happy, sad, surprise, calmness}. Correspondingly, in the above emotion recognition algorithm, we also identified these seven emotions. Naoyuki KUBOTA and Yusuke NOJIMA have proposed a pet robot with an emotional model [28]. In their model, external stimuli are considered to be instantaneous, and mood [29] is determined by the value of mood before stimulation and the value of stimuli. The value of mood does not change between the two stimuli. The advantage of this model is that it takes into account the influence of the external environment on emotions , and through iteration, realizes the long-term changes of emotions, and to some extent reflects the "memory" function of robot emotions. In a study by Sakmongkon Chumkamon [27], they simulated changes in the effects of various external stimuli on robots by simulating the changes in animal dopamine. The current emotion of the robot is calculated by combining the curves of various external stimuli and calculating them. The advantage of this model is that it considers the variability of the influence of external stimuli on the emotions of the robot, and distinguishes a variety of different stimuli, which can better express the emotions of the robot and be closer to the behavior of the animal. But its ability to simulate long-term emotional memory is not strong. Our model has been synthesized and improved on the basis of the above two models.

This paper uses a probabilistic model to represent possible changes in animal emotions, which is shown in Fig. 14. For example, if the current robot emotion is $i$, then the vector $\{p_{i1}, \cdots, p_{ij}, \cdots, p_{in}\}$ represents the probability of switching from emotion $i$ to all the $n$ emotions. The robot will periodically update this vector value, and after a period of time, select the $k$ transitions with the highest probability, and then randomly convert to an emotion and generate a new vector according to its probability. And $k$ is the threshold to prevent a very low probability transition from occurring.

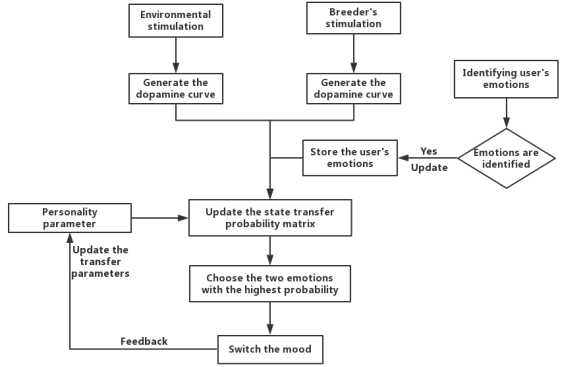

Fig. 14. Emotional interaction model architecture.

The calculation process of the conversion probability $p_{ij}$ of emotion i to j is

$$M_{ij} = w_{ij} * (\alpha_1 * (S_1 + S_3) + \alpha_2 * (S_2 + S_4)) + W_j \quad (8)$$
$$P_{ij} = M_{ij}/\sum_{k=1}^{n} M_{ik} \quad (9)$$

Where $w_{ij}$ is a statistical value converted from emotion i to emotion j, which will be described in detail below. $\alpha_1$ and $\alpha_2$ represent the positive and negative correlations of emotional influence factors, and the value is -1 or +1. $S_1 \sim S_4$ respectively indicate positive stimulation of the environment, negative stimulation of the environment, positive stimulation of the breeder, and negative stimulation of the breeder. $S_1$ and $S_2$ are determined by the comfort of the environment. The environmental comfort E is calculated by calculating the values obtained from the various sensors.

$$E = \alpha_1 * e_1 + \alpha_2 * e_2 + \cdots + \alpha_n * e_n \quad (10)$$

Where $\alpha$ represents the weight of the input values of each sensor, and $e$ represents the input values of each sensor.

When the comfort level is higher than our given threshold C, the negative stimulus $S_2$ of the environment is zero, and the positive stimulus $S_1$ is E-C. Conversely, when the comfort level is below the threshold, the environment positive stimulus $S_1$ is zero, and the environmental negative stimulus $S_2$ is C-E.

$S_3$ and $S_1$ are determined by the interaction between the pet breeder and the pet. When the pet gets the favorite virtual props, $S_4$ is zero, $S_3$ becomes positive, and the value is determined by the virtual props. Similarly, when the pet gets disliked the virtual props, $S_4$ becomes a positive value and $S_3$ becomes zero.

When calculating $M_{ij}$, we will also choose different values of $\alpha$ according to the kind of emotion j. We divide emotions into two broad classes, namely, positive class {happy, surprise, netural} and negative class {sad, anger, disgust, fear}. For example, when j is in a negative class, the negative external influence will increase its formation, while the positive influence will slow it down, so $\alpha_1$ takes -1 and $\alpha_2$ takes +1.

After calculating $S_1 \sim S_4$, the value of the emotional impact will be updated separately, which will gradually decrease with time. Here we follow the dopamine simulation model proposed in the literature [17].

$$\ddot{y}_{(t)} + 2\xi\omega_n\dot{y}_{(t)} + \omega_n^2 y_{(t)} = \omega_n^2 u_{(t)} \quad (11)$$
$$y_{(t)} = e^{-t/Tc} \cdot y_{peak} \quad (12)$$

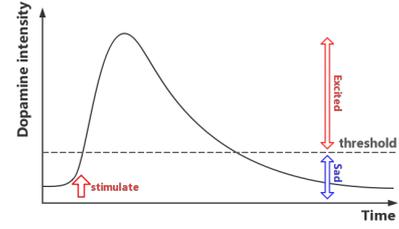

Fig. 15. Simulated dopamine change curve.

The vector $W = \{W_1, \cdots, W_j, \cdots, W_n\}$ is the preset pet personality value, which will be determined by the pet breeder himself, for example, if the breeder prefers that his pet is more optimistic, that means the value of $W_i$ that switches to optimistic emotions will be bigger. In addition, W is not a constant, and it will be dynamically updated after each emotional transition. For example, if the emotion of the robot is switched to j, the value of $W_j$ will increase. We hope that the character of the robot will gradually stabilize during the long-term interaction with the outside world.

$$W_j(n + 1) = (1 + \beta)W_j(n) \quad (13)$$

Where $\beta$ is the growth coefficient, which determines the speed at which the robot character tends to stabilize. Thus, over time, the proportion of W in the emotional transformation of robots will gradually increase.

Below we will discuss the calculation of $W_{ij}$. The value of $W_{ij}$ is determined not only by the current state of the robot, but also by the emotions of the user, which is in line with our pre-defined robotic functions that can sense emotions and overcome autism. In order to determine the value of $W_{ij}$, we conducted a questionnaire among 1,000 people. An example of the contents of the questionnaire is as follows.

When your emotions are happy, if you see your good friend is sad, what will your emotions be?

1 happy, 2sad, 3anger, 4disgust, 5 fear, 6surprise, 7neutral

One form of the histogram of the results we get is in Fig.16.

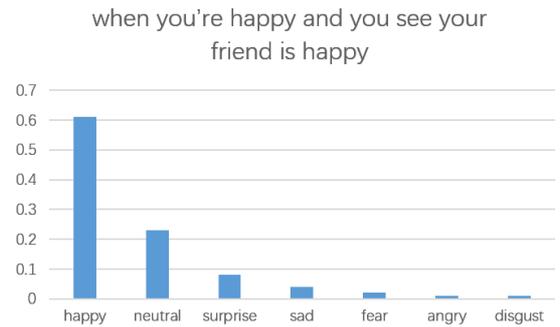

Fig. 16. Histogram of the questionnaire results

## V. EXPERIMENT AND ANALYSIS

In order to verify the effect of the "Tom-Talker" social mechanism on urban autism in the specific application process, this paper designed an experiment, which is in the "Tom-Talk" social platform and the "Tom" pet robot to see whether they are able to achieve the intended purpose.

### A. "Tom-Talker" Social Platform Experiment

In order to validate our social model, we randomly selected 400 people in a certain area to participate in our experiment.

We randomly divided them equally into two groups, one using our social model and the other as a comparison, and in order to control the cost of the experiment, we chose an electronic pet with the same feeding functions as "TOM". This way the user will be able to feed the pet by obtaining virtual props on our platform. We separately calculated the following indicators of the experimenters before and after the experiment.

Weekly social time is time each person surveyed spends for social activities on a weekly basis. It can effectively reflect the amount of social activity of respondents. The social circle size is the number of people in the social circle of the surveyed people, which can effectively reflect the social extent of them. The averages of these two indicators for each group are shown in Fig. 17 and Fig. 18.

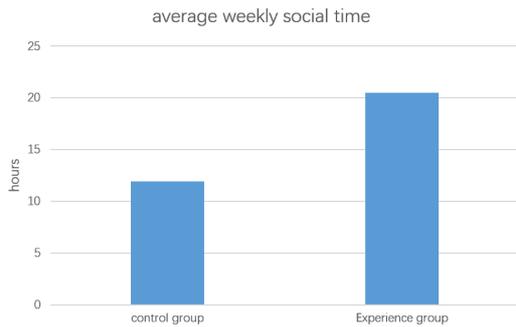

Fig. 17. Average weekly social time comparison chart

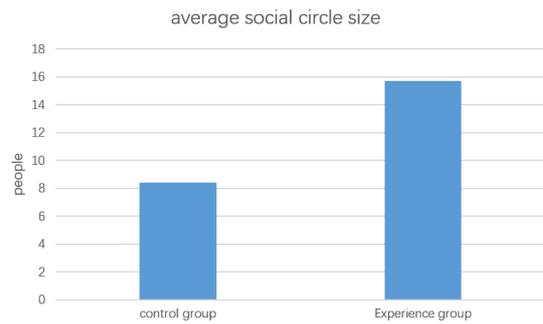

Fig. 18. Social circle size comparison chart

In addition, we used the UCLA Loneliness Scale [18] proposed by the predecessors to investigate the subjective loneliness of the respondents. The results are shown in Fig. 19 and Fig.20.

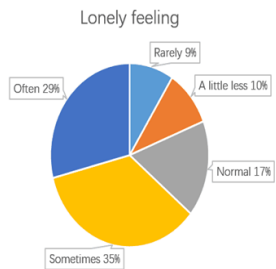

Fig. 19. The control group's loneliness survey pie chart

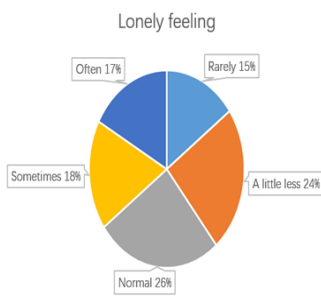

Fig. 20. The test group's loneliness survey pie chart

The experimental results show that although we have not used a more attractive physical pet robot, only the benefits of its social model have been very significant.

### B. Pet Robot Emotional Interaction Experiment

For our emotional interaction model, we chose 20 people to experiment and equip each person with a "Tom" robot to Further verify our model by observing the interaction between the experimenter and the robot.

User satisfaction refers to whether the user is satisfied with the emotional model of the robot, such as whether it can correctly recognize the emotion and can respond accordingly to alleviate the user's loneliness reasonably. We divided it into five levels: {very dissatisfied, dissatisfied, Okay, satisfied, very satisfied}, and conducted a questionnaire survey on users. The result is shown in Fig. 21.

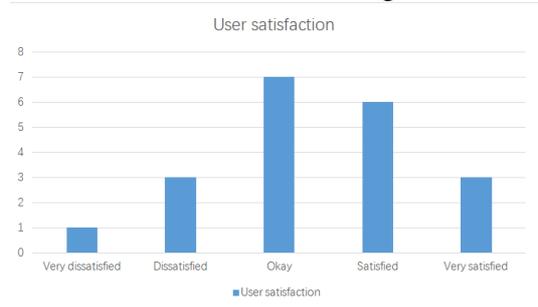

Fig. 21. User satisfaction survey histogram

The emotion recognition situation refers to the recognition rate of the user's emotions by the robot. Here we show the confusion matrix of our robot's emotion recognition, which is shown in Fig.22.

| | happy | neutral | surprise | sad | fear | angry | disgust |
|---|---|---|---|---|---|---|---|
| happy | 0.83 | 0.10 | 0.11 | 0.07 | 0.06 | 0.06 | 0.06 |
| neutral | 0.07 | 0.63 | 0.08 | 0.19 | 0.10 | 0.12 | 0.05 |
| surprise | 0.02 | 0.04 | 0.65 | 0.02 | 0.12 | 0.02 | 0.11 |
| sad | 0.02 | 0.08 | 0.04 | 0.51 | 0.14 | 0.14 | 0.02 |
| fear | 0.02 | 0.06 | 0.08 | 0.07 | 0.42 | 0.08 | 0.03 |
| angry | 0.03 | 0.07 | 0.03 | 0.12 | 0.13 | 0.59 | 0.21 |
| disgust | 0.01 | 0.02 | 0.01 | 0.02 | 0.03 | 0.01 | 0.52 |

Fig. 22. Emotion recognition confusion matrix

The results show that our emotional model can well recognize the user's emotions and respond well to the user's emotions.

## VI. CONCLUSION

This article introduces a series of social issues such as urban autism generated by the rapid development of network information, and proposes a solution to this problem by using the "Tom" pet robot.

This paper mainly designs a new intelligent pet robot, a pet social mechanism, and an emotional interaction algorithm based on the robot. In terms of social mechanism, this paper designs a "Tom-Talker" robot social mechanism to establish connections between users ,robots and social circles based on a reward mechanism and friend recommendation algorithm, to help users get out of the house and socialize with others. In terms of robotic emotional interaction, this paper designs a pet robot emotional interaction algorithm that can recognize users' emotions and simulate animals' emotions, so as to establish emotional communication between pet robots and users.

Finally, this paper designs experiments and analyzes the experimental results. Experiments show that pet robots have a good effect in solving urban autism problems. In addition, the robot also has good stability, simple structure, rich functions and direct effects.

ACKNOWLEDGMENT

The corresponding author is Zhao Hongwei and Xingqian Li. The authors are grateful to the anonymous reviewers for their insightful comments which have certainly improved this paper.

This work was supported in part by the Jilin provincial science and technology innovation project 20190302026GX and State Key Laboratory of Applied Optics Open Fund Project 20173660.

Compliance with Ethical Standards:

Funding: This study was funded by 20190302026GX and 20173660.

Conflict of Interest: Author Hongwei Zhao and Xingqian Li has received research grants from Company Jilin province development and Reform Commission Special industrial innovation and State Key Laboratory of Applied Optics. The authors declare that they have no conflict of interest.